\documentstyle[12pt,epsf,axodraw]{article} \textwidth 16cm
\newcommand{\beq}[1]{
\begin{equation}\label{#1}}
\newcommand{\eeq}{\end{equation}}
\newcommand{\bea}[1]{
\begin{eqnarray}\label{#1}}
\newcommand{\eea}{\end{eqnarray}}
\begin{document}

\begin{titlepage}

\begin{center}
{\LARGE \bf Electroproduction of two light vector mesons in next-to-leading
BFKL: study of systematic effects}

\vspace{1cm}

{\sc D.Yu.~Ivanov}${}^{1}$ and
{\sc A.~Papa}${}^{2}$
\\[0.5cm]
\vspace*{0.1cm}
${}^1$ {\it
Sobolev Institute of Mathematics, 630090 Novosibirsk, Russia
                       } \\[0.2cm]
\vspace*{0.1cm} ${}^2$ {\it Dipartimento di Fisica, Universit\`a
della Calabria \\
and Istituto Nazionale di Fisica Nucleare, Gruppo collegato di Cosenza \\
I-87036 Arcavacata di Rende, Cosenza, Italy
                       } \\[1.0cm]

\vspace*{0.5cm}

\centerline{\large \em \today}

\vskip2cm
{\bf Abstract\\[10pt]} \parbox[t]{\textwidth}
{The forward electroproduction of two light vector mesons is the first example of a
collision process between strongly interacting colorless particles for
which the amplitude can be written completely within perturbative QCD in the Regge
limit with next-to-leading accuracy. In a previous paper we have given a
numerical determination of the amplitude in the case of equal photon virtualities
by using a definite representation for the amplitude and a definite optimization
method for the perturbative series. Here we estimate the systematic uncertainty
of our previous determination, by considering a different representation
of the amplitude and different optimization methods of the perturbative series.
Moreover, we compare our result for the differential cross section at
the minimum $|t|$ with a different approach, based on collinear
kernel improvement.}

\vskip1cm
\end{center}

\vspace*{1cm}

\end{titlepage}

\section{Introduction}

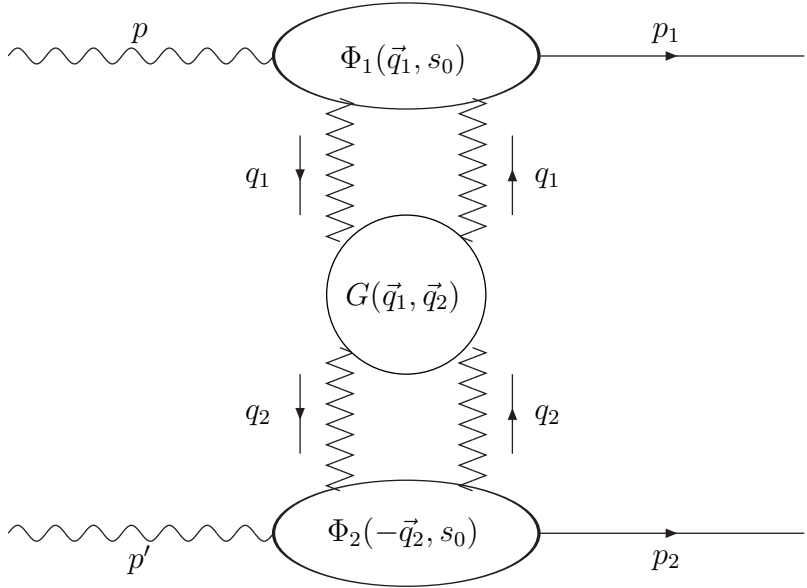
\begin{figure}[tb]
\centering
\setlength{\unitlength}{0.35mm}
\begin{picture}(300,200)(0,0)

\Photon(0,190)(100,190){3}{7}
\ArrowLine(200,190)(300,190)
\Text(50,200)[c]{$p$}
\Text(250,200)[c]{$p_1$}
\Text(150,190)[]{$\Phi_1(\vec q_1, s_0)$}
\Oval(150,190)(20,50)(0)

\ZigZag(125,174)(125,120){5}{7}
\ZigZag(175,174)(175,120){5}{7}
\ZigZag(125,26)(125,80){5}{7}
\ZigZag(175,26)(175,80){5}{7}

\ArrowLine(110,160)(110,130)
\ArrowLine(190,130)(190,160)
\ArrowLine(110,70)(110,40)
\ArrowLine(190,40)(190,70)

\Text(100,145)[r]{$q_1$}
\Text(200,145)[l]{$q_1$}
\Text(100,55)[r]{$q_2$}
\Text(200,55)[l]{$q_2$}

\GCirc(150,100){30}{1}
\Text(150,100)[]{$G(\vec q_1,\vec q_2)$}

\Photon(0,10)(100,10){3}{7}
\ArrowLine(200,10)(300,10)
\Text(50,0)[c]{$p'$}
\Text(250,0)[c]{$p_2$}
\Text(150,10)[]{$\Phi_2(-\vec q_2,s_0)$}
\Oval(150,10)(20,50)(0)

\end{picture}

\caption[]{Schematic representation of the amplitude for the $\gamma^*(p)\,
\gamma^*(p') \to V(p_1)\, V(p_2)$ scattering.}
\label{fig:BFKL}
\end{figure}

The BFKL approach~\cite{BFKL} to strong interactions is expected to describe
collision processes with a large center-of-mass energy and with a ``hard''
enough scale to permit the use of perturbative expansion in the
strong coupling $\alpha_s$. In this approach, both in the leading logarithmic
approximation (LLA), which means resummation of leading energy logarithms,
all terms $(\alpha_s\ln(s))^n$, and in the next-to-leading approximation (NLA),
which means resummation of all terms $\alpha_s(\alpha_s\ln(s))^n$, the
(imaginary part of the) amplitude for a large-$s$ hard collision process can be
written as the convolution of the Green's function of two interacting Reggeized
gluons with the impact factors of the colliding particles (see, for example,
Fig.~\ref{fig:BFKL}).

The Green's function is determined through the BFKL equation. The NLA singlet
kernel of the BFKL equation has been achieved in the forward
case~\cite{NLA-kernel}, after the long program of calculation of the NLA
corrections~\cite{NLA-corrections} (for a review, see Ref.~\cite{news}).
For the non-forward case the ingredients to the NLA BFKL kernel are known since
a few years for the color octet representation in the
$t$-channel~\cite{NLA-corrections-nf}. This color representation is very
important for the check of consistency of the $s$-channel unitarity with
the gluon Reggeization, i.e. for the ``bootstrap''~\cite{bootstrap}.
Recently the non-forward NLA BFKL kernel has been derived also
in the singlet color representation, i.e. in the Pomeron channel, relevant
for physical applications~\cite{FF05}.

On the side of impact factors, however, only a limited knowledge is available.
Impact factors have been calculated with NLA accuracy for colliding
partons~\cite{partonIF} and for forward jet production~\cite{BCV03}.
The most important impact factor for the BFKL phenomenology, the
$\gamma^* \to \gamma^*$ impact factor, is calling for a rather long
calculation, although it seems to be close to completion now~\cite{gammaIF}.
The only available colorless impact factor is presently the one for the
forward transition from a virtual photon $\gamma^*$ to a light neutral vector meson
$V=\rho^0, \omega, \phi$, obtained in Ref.~\cite{IKP04}. This impact factor
can be used, together with the NLA BFKL Green's function, to build
completely within perturbative QCD and with NLA accuracy the
amplitude of the $\gamma^* \gamma^* \to V V$ reaction. This amplitude provides us
with an ideal theoretical laboratory for the investigation of several open
questions in the BFKL approach and for the comparison with different approaches.

Such investigation was started in Ref.~\cite{IP06}, where it was first
of all shown how the $\gamma^* \to V$ impact factors and the BFKL Green's function
can be put together to build up the NLA forward amplitude of the $\gamma^* \gamma^*
\to V V$ process in the $\overline {\mbox{MS}}$ scheme and that a convenient series
representation for this amplitude can be determined. Then, in the case of
equal photon virtualities, i.e. in the so-called ``pure'' BFKL regime, a numerical
study was carried out which led to conclude that the NLA corrections are large and
of opposite sign with respect to the leading order and that they are dominated,
at the lower energies, by the NLA correction from impact factors. However,
this fact did not prevent us from achieving a smooth behavior of the
(imaginary part of the) amplitude with the energy, by optimizing the choice of
the energy scale $s_0$ in the BFKL approach and the renormalization scale $\mu_R$
which appear in subleading terms. The optimization method adopted there was an
adaptation of the ``principle of minimum sensitivity'' (PMS)~\cite{Stevenson}
to the case where two energy parameters are present.

The aim of this paper is to estimate the amount of the systematic effects in the
determination of Ref.~\cite{IP06}, by exploring the two main sources:
the choice of the representation of the amplitude and the choice of the optimization
method. Concerning the first effect, in this paper we compare the series representation
of the amplitude with another representation, equivalent to the previous
with NLA accuracy, where almost all the NLA corrections coming from the kernel are
exponentiated. As for the second effect, we compare here the PMS optimization method
with two other well-known methods of optimization of the perturbative series, namely
the fast apparent convergence (FAC) method~\cite{Grun} and the
Brodsky-Lepage-Mackenzie (BLM) method~\cite{BLM}.

It would be quite interesting to apply to the amplitude under consideration
here the improvement of the NLA BFKL kernel as a consequence of the analysis of
collinear singularities of the NLA corrections and by the account of
further collinear terms beyond NLA~\cite{Sal98,Ciafaloni,
Altarelli,Thorne,agustin,Peschanski}. As pointed out in Ref.~\cite{IP06},
the strategy of collinear improvement has something in common with ours,
in the sense that it is also inspired by renormalization-group invariance
and it also leads to the addition of terms beyond the NLA. Work is in
progress in this direction.

In this paper, however, we present a comparison with an approach by another
research group based on the collinear improvement of the kernel. In particular,
we compare our determinations for the differential cross section at the minimum
$|t|$ of the $\gamma^* \gamma^* \to V V$ process for two values
of the common photon virtuality with the results of Ref.~\cite{EPSW1}, where
the same process has been considered using some version of a collinearly improved
kernel.

This paper is organized as follows: in the next Section we briefly recall
the relevant notation and give the two representations of the amplitude, series and
``exponentiated'', to be considered in the following; in Section~3 we recall
the strategies of the three optimization methods considered and perform the
numerical comparisons; in Section~4 we compare our differential cross section with
that of Ref.~\cite{EPSW1}; in Section~5 we draw our conclusions.

\section{Representations of the NLA amplitude}

The process under consideration is the production of two light vector
mesons ($V=\rho^0, \omega, \phi$) in the collision of two virtual photons,
\beq{process}
\gamma^*(p) \: \gamma^*(p')\to V(p_1) \:V(p_2) \;.
\eeq
Here, neglecting the meson mass $m_V$, $p_1$ and $p_2$ are taken as Sudakov vectors satisfying $p_1^2=p_2^2=0$ and
$2(p_1 p_2)=s$; the virtual photon momenta are instead
\beq{kinphoton}
p =\alpha p_1-\frac{Q_1^2}{\alpha s} p_2 \;, \hspace{2cm}
p'=\alpha^\prime p_2-\frac{Q_2^2}{\alpha^\prime s} p_1 \;,
\eeq
so that the photon virtualities turn to be $p^2=-Q_1^2$ and $(p')^2=-Q_2^2$.
We consider the kinematics when
\beq{kin}
s\gg Q^2_{1,2}\gg \Lambda^2_{QCD} \, ,
\eeq
and
\beq{alphas}
\alpha=1+\frac{Q_2^2}{s}+{\cal O}(s^{-2})\, , \quad
\alpha^\prime =1+\frac{Q_1^2}{s}+{\cal O}(s^{-2})\, .
\eeq
In this case vector mesons are produced by longitudinally polarized photons in
the longitudinally polarized state~\cite{IKP04}. Other helicity amplitudes are
power suppressed, with a suppression factor $\sim m_V/Q_{1,2}$.
We will discuss here the amplitude of the forward scattering, i.e.
when the transverse momenta of produced $V$ mesons are zero or
when the variable $t=(p_1-p)^2$ takes its maximal value $t_0=-Q_1^2Q_2^2/s+{\cal
O}(s^{-2})$.

The forward amplitude in the BFKL approach may be presented as follows
\beq{imA}
{\cal I}m_s\left( {\cal A} \right)=\frac{s}{(2\pi)^2}\int\frac{d^2\vec q_1}{\vec
q_1^{\,\, 2}}\Phi_1(\vec q_1,s_0)\int
\frac{d^2\vec q_2}{\vec q_2^{\,\,2}} \Phi_2(-\vec q_2,s_0)
\int\limits^{\delta +i\infty}_{\delta
-i\infty}\frac{d\omega}{2\pi i}\left(\frac{s}{s_0}\right)^\omega
G_\omega (\vec q_1, \vec q_2)\, .
\eeq
This representation for the amplitude is valid with NLA accuracy.
Here $\Phi_{1}(\vec q_1,s_0)$ and $\Phi_{2}(-\vec q_2,s_0)$
are the impact factors describing the transitions $\gamma^*(p)\to V(p_1)$
and $\gamma^*(p')\to V(p_2)$, respectively.
The Green's function in (\ref{imA}) obeys the BFKL equation
\beq{Green}
\delta^2(\vec q_1-\vec q_2)=\omega \, G_\omega (\vec q_1, \vec q_2)-
\int d^2 \vec q \, K(\vec q_1,\vec q)\, G_\omega (\vec q, \vec q_2) \;,
\eeq
where $K(\vec q_1,\vec q_2)$ is the BFKL kernel.
The scale $s_0$ is artificial. It is introduced in the BFKL approach at the time
to perform the Mellin transform from the $s$-space to the complex angular
momentum plane and must disappear in the full expression for the amplitude
at each fixed order of approximation. Using the result for the meson
NLA impact factor such cancellation was demonstrated explicitly in
Ref.~\cite{IKP04} for the process in question.

The impact factors are also presented as an expansion in $\alpha_s$
\beq{impE}
\Phi_{1,2}(\vec q)= \alpha_s \,
D_{1,2}\left[C^{(0)}_{1,2}(\vec q^{\,\, 2})+\bar\alpha_s
C^{(1)}_{1,2}(\vec
q^{\,\, 2})\right] \, , \quad D_{1,2}=-\frac{4\pi e_q  f_V}{N_c Q_{1,2}}
\sqrt{N_c^2-1}\, ,
\eeq
where $f_V$ is the meson dimensional coupling constant ($f_{\rho}\approx
200\, \rm{ MeV}$) and $e_q$ should be replaced by $e/\sqrt{2}$, $e/(3\sqrt{2})$
and $-e/3$ for the case of $\rho^0$, $\omega$ and $\phi$ meson production,
respectively.

In the collinear factorization approach the meson transition impact factor
is given as a convolution of the hard scattering amplitude for the
production of a collinear quark--antiquark pair with the meson distribution
amplitude (DA). The integration variable in this convolution is the fraction $z$
of the meson momentum carried by the quark ($\bar z\equiv 1-z$ is
the momentum fraction carried by the antiquark):
\beq{imps1}
C^{(0)}_{1,2}(\vec q^{\,\, 2})=\int\limits^1_0 dz \,
\frac{\vec q^{\,\, 2}}{\vec q^{\,\, 2}+z \bar zQ_{1,2}^2}\phi_\parallel (z)
\, .
\eeq

The NLA correction to the hard scattering amplitude, for a photon with virtuality
equal to $Q^2$, is defined as follows
\beq{imps2}
C^{(1)}(\vec q^{\,\, 2})=\frac{1}{4 N_c}\int\limits^1_0 dz \,
\frac{\vec q^{\,\, 2}}{\vec q^{\,\, 2}+z \bar zQ^2}[\tau(z)+\tau(1-z)]
\phi_\parallel (z)
\, ,
\eeq
with $\tau(z)$ given in the Eq.~(75) of Ref.~\cite{IKP04}.
$C^{(1)}_{1,2}(\vec q^{\,\, 2})$ are given by the previous expression with
$Q^2$ replaced everywhere in the integrand by $Q^2_1$ and $Q^2_2$,
respectively. We use the distribution amplitude in its asymptotic form
$\phi^{as}_\parallel(z)=6z(1-z)$. The main reason for this choice is
simplicity. The other point is that for the case of
equal photon virtualities the typical values of the Reggeon momenta are
$\vec q^{\,\, 2}\sim Q^2$. In this case the integrands in
Eqs.~(\ref{imps1}) and~(\ref{imps2}) are smooth functions of $z$ and,
consequently, the amplitude is not very sensitive to the shape of the
meson DA.

In Ref.~\cite{IP06} the NLA forward amplitude has been written
as a spectral decomposition on the basis of eigenfunctions of the
LLA BFKL kernel:
\[
\frac{{\cal I}m_s\left( {\cal A} \right)}{D_1D_2}=\frac{s}{(2\pi)^2}
\int\limits^{+\infty}_{-\infty}
d\nu \left(\frac{s}{s_0}\right)^{\bar \alpha_s(\mu_R) \chi(\nu)}
\alpha_s^2(\mu_R) c_1(\nu)c_2(\nu)\left[1+\bar \alpha_s(\mu_R)
\left(\frac{c^{(1)}_1(\nu)}{c_1(\nu)}
+\frac{c^{(1)}_2(\nu)}{c_2(\nu)}\right)
\right.
\]
\beq{amplnla}
\left.
+\bar \alpha_s^2(\mu_R)\ln\left(\frac{s}{s_0}\right)
\left(\bar
\chi(\nu)+\frac{\beta_0}{8N_c}\chi(\nu)\left[-\chi(\nu)+\frac{10}{3}
+i\frac{d\ln(\frac{c_1(\nu)}{c_2(\nu)})}{d\nu}+2\ln(\mu_R^2)\right]
\right)\right] \; .
\eeq
Let us briefly recall the definition of the objects entering this
expression:
\beq{baral}
{\bar \alpha_s}=\frac{\alpha_s N_c}{\pi} \;,
\eeq
with $N_c$ the number of colors;
\beq{chi}
\chi (\nu)=2\psi(1)-\psi\left(\frac{1}{2}+i\nu\right)-\psi\left(\frac{1}{2}
-i\nu\right)\, ,
\eeq
\beq{imp1}
c_1(\nu)=\int d^2\vec q \,\, C_1^{(0)}(\vec q^{\, 2})
\frac{\left(\vec q^{\, 2}\right)^{i\nu-\frac{3}{2}}}{\pi \sqrt{2}}
\, ,\;\;\;\;\;
c_2(\nu)=\int d^2\vec q \,\, C_2^{(0)}(\vec q^{\, 2})
\frac{\left(\vec q^{\, 2}\right)^{-i\nu-\frac{3}{2}}}{\pi \sqrt{2}} \, ,
\eeq
and similar equations for $c_1^{(1)}(\nu)$ and $c_2^{(1)}(\nu)$
from the NLA corrections to the impact factors, $C_1^{(1)}(\vec
q^{\,\, 2})$ and $C_2^{(1)}(\vec q^{\,\, 2})$;
\bea{chibar}
\bar \chi(\nu)\,&=&\,-\frac{1}{4}\left[\frac{\pi^2-4}{3}\chi(\nu)-6\zeta(3)-
\chi^{\prime\prime}(\nu)-\frac{\pi^3}{\cosh(\pi\nu)}
\right.
\nonumber \\
&+& \left.
\frac{\pi^2\sinh(\pi\nu)}{2\,\nu\, \cosh^2(\pi\nu)}
\left(
3+\left(1+\frac{n_f}{N_c^3}\right)\frac{11+12\nu^2}{16(1+\nu^2)}
\right)
+\,4\,\phi(\nu)
\right] \, ,
\eea
\beq{phi}
\phi(\nu)\,=\,2\int\limits_0^1dx\,\frac{\cos(\nu\ln(x))}{(1+x)\sqrt{x}}
\left[\frac{\pi^2}{6}-\mbox{Li}_2(x)\right]\, , \;\;\;\;\;
\mbox{Li}_2(x)=-\int\limits_0^xdt\,\frac{\ln(1-t)}{t} \, .
\eeq

Using Eq.~(\ref{amplnla}) we construct the following representation for the
amplitude
\bea{series}
\frac{Q_1Q_2}{D_1 D_2} \frac{{\cal I}m_s ({\cal A}_{\mathrm series})}{s} &=&
\frac{1}{(2\pi)^2}  \alpha_s(\mu_R)^2 \label{honest_NLA} \\
& \times &
\biggl[ b_0
+\sum_{n=1}^{\infty}\bar \alpha_s(\mu_R)^n   \, b_n \,
\biggl(\ln\left(\frac{s}{s_0}\right)^n   +
d_n(s_0,\mu_R)\ln\left(\frac{s}{s_0}\right)^{n-1}     \biggr)
\biggr]\;, \nonumber
\eea
where the coefficients
\beq{bs}
\frac{b_n}{Q_1Q_2}=\int\limits^{+\infty}_{-\infty}d\nu \,  c_1(\nu)c_2(\nu)
\frac{\chi^n(\nu)}{n!} \, ,
\eeq
are determined by the kernel and the impact factors in LLA.
The coefficients
\[
d_n=n\ln\left(\frac{s_0}{Q_1Q_2}\right)+\frac{\beta_0}{4N_c}
\left(
(n+1)\frac{b_{n-1}}{b_n}\ln\left(\frac{\mu_R^2}{Q_1Q_2}\right)
-\frac{n(n-1)}{2} \right.
\]
\beq{ds}
\left.
+ \frac{Q_1Q_2}{b_n}\int\limits^{+\infty}_{-\infty}d\nu \, (n+1)f(\nu)
c_1(\nu)c_2(\nu)
\frac{\chi^{n-1}(\nu)}{(n-1)!}\right)
\eeq
\[
+\frac{Q_1Q_2}{b_n}\left(
\int\limits^{+\infty}_{-\infty}d\nu\, c_1(\nu)c_2(\nu)
\frac{\chi^{n-1}(\nu)}{(n-1)!}\left[
\frac{\bar c^{(1)}_{1}(\nu)}{c_{1}(\nu)}+\frac{\bar
c^{(1)}_{2}(\nu)}{c_{2}(\nu)}
 +(n-1)\frac{\bar \chi(\nu)}{\chi(\nu)}\right]
\right)
\]
are determined by the NLA corrections to the kernel and to the impact
factors. Here $\bar c^{(1)}_{1,2}$ are determined according to
the definitions~(\ref{imp1}) from
\beq{sepa}
\bar C^{(1)}(\vec q^{\,\,2})= C^{(1)}(\vec q^{\,\,2})
\eeq
\[
-\int\limits^1_0 dz \,
\frac{\vec q^{\,\, 2}}{\vec q^{\,\, 2}+z \bar zQ^2}\phi_\parallel (z)
\left[
\frac{1}{4}\ln\left(\frac{s_0}{Q^2}\right)\ln\left(\frac{(\alpha+z\bar
z)^4}{\alpha^2 z^2\bar z^2}\right)+\frac{\beta_0}{4N_c}\left(
\ln\left(\frac{\mu_R^2}{Q^2}\right)+\frac{5}{3}-\ln(\alpha)\right)
\right]\;.
\]
Moreover, we use the notation
\beq{fv}
f(\nu)=\frac{5}{3}+\psi(3+2i\nu)+\psi(3-2i\nu)-\psi\left(\frac{3}{2}+i\nu\right)
-\psi\left(\frac{3}{2}-i\nu\right) \, .
\eeq

One should stress that both representations of the amplitude~(\ref{series})
and~(\ref{amplnla}) are equivalent with NLA accuracy, since they differ only by
next-to-NLA (NNLA) terms. The series representation~(\ref{series}) is
a natural choice, since it includes in some sense the minimal amount of
NNLA contributions; moreover, its form is the closest one to the initial
goal of the BFKL approach, i.e. to sum selected contributions in the
perturbative series.

Actually there exist infinitely many possibilities to write a NLA
amplitude. The other possibility considered here is to exponentiate the bulk
of the kernel NLA corrections
\[
\frac{{\cal I}m_s\left( {\cal A}_{\mathrm exp}\right)}{D_1D_2}
=\frac{s}{(2\pi)^2}
\int\limits^{+\infty}_{-\infty}
d\nu \left(\frac{s}{s_0}\right)^{\bar \alpha_s(\mu_R)
\chi(\nu)+\bar \alpha_s^2(\mu_R)
\left(
\bar
\chi(\nu)+\frac{\beta_0}{8N_c}\chi(\nu)\left[-\chi(\nu)+\frac{10}{3}
\right]
\right)}
\alpha_s^2(\mu_R) c_1(\nu)c_2(\nu)
\]
\beq{amplnlaE}
\times\! \left[1+\bar \alpha_s(\mu_R)
\left(\frac{c^{(1)}_1(\nu)}{c_1(\nu)}
+\frac{c^{(1)}_2(\nu)}{c_2(\nu)}\right)
+\bar \alpha_s^2(\mu_R)\ln\left(\frac{s}{s_0}\right)
\frac{\beta_0}{8N_c}\chi(\nu)\left(
i\frac{d\ln(\frac{c_1(\nu)}{c_2(\nu)})}{d\nu}+2\ln(\mu_R^2)
\right)\right].
\eeq
This form of the NLA amplitude was used in~\cite{KIM1} (see also~\cite{KIM2}),
without account of the last two terms in the second line of (\ref{amplnlaE}),
for the analysis of the total $\gamma^*\gamma^*$ cross section. We will
refer in the following to this representation simply as ``exponentiated''
amplitude.

It is easily seen from Eqs.~(\ref{series})-(\ref{fv}) that the amplitude is
independent in the NLA from the choice of energy and strong coupling
scales~\cite{IP06}.

For the purposes of our analysis of systematic effects, 
it could be acceptable also to use an amplitude where the last two terms in 
the squared bracket of the integrand in the R.H.S. of Eq.~(\ref{amplnlaE}) are 
exponentiated. Indeed, we performed a numerical analysis also in this case 
obtaining results which are in fair agreement with our findings below. 
Nevertheless, we prefer not to exponentiate these terms for the following reason. 
By exponentiation we mean to transfer an effect from the kernel to the Green's 
function, which means to account for the corresponding effect "to all orders".
The last term in Eq.~(\ref{amplnlaE}) originates from the scale non-invariant part of the 
NLA kernel, i.e. from the running of the coupling, which leads in the Mellin 
space to the derivative term, see Eq.~(25) in Ref.~\cite{IP06}.
It is known~\cite{Armesto:1998gt}, that an exact account of the derivative term 
in the BFKL equation leads to a radical change of both the spectrum and the
eigenfunctions of the NLA BFKL kernel. Another point is that
BFKL approach itself (where an amplitude is a convolution of the 
Green's function and the impact factors) is valid only within NLA. In 
higher orders one should take into account additional contributions, 
related, for instance, with the transition of two to four Reggeized gluons 
propagating in the $t$-channel. In this situation we decided to leave
the last two terms in Eq.~(\ref{amplnlaE}) unexponentiated.

Another interesting representation for the BFKL NLA amplitude
appears if one treats the energy scale parameters dynamically,
allowing $s_0$ being a function of the Reggeons transverse
momenta. Such change of energy scale leads to the corresponding
modification of the impact factors, and even of the kernel of BFKL
equation in the case if the energy scale does not factorize as the
product of two functions of $\vec q_1$ and $\vec q_2$, see~\cite{Fadin:1998sg}.
However numerical implementation of such dynamical energy scale scheme
would require the knowledge of NLA BFKL Green's function in the momentum 
representation, $G(\vec q_1,\vec q_2)$, and its subsequent integration 
with the impact factors. Such a study could be done with the methods developed 
in Ref.~\cite{Andersen:2003wy}.

\section{Numerical results}

In Ref.~\cite{IP06} we have presented some numerical results for the
amplitude given in Eq.~(\ref{series}) for the $Q_1=Q_2\equiv Q$ kinematics,
i.e. in the ``pure'' BFKL regime. We truncated the series in the
R.H.S. of Eq.~(\ref{series}) to $n=20$, after having verified that
this procedure gives a very good approximation of the infinite sum
for the $Y$ values $Y\leq 10$ and used the two--loop running coupling
corresponding to the value $\alpha_s(M_Z)=0.12$.

We obtained there the following results for the $b_n$ and $d_n$ coefficients
for $n_f=5$ and $s_0=Q^2=\mu_R^2$:
\beq{coe}
\begin{array}{lllll}
b_0=17.0664  & b_1=34.5920   & b_2=40.7609  & b_3=33.0618   &
b_4=20.7467  \\
             & b_5=10.5698  & b_6=4.54792 & b_7=1.69128  &
b_8=0.554475\\
& & & & \\
& d_1=-3.71087 & d_2=-11.3057 & d_3=-23.3879 & d_4=-39.1123 \\
& d_5=-59.207 & d_6=-83.0365 & d_7=-111.151 & d_8=-143.06 \;, \\
\end{array}
\eeq
the main contribution to the $d_n$ coefficients originating from
the NLA corrections to the impact factors for $n\leq 3$.

\begin{figure}[tb]
\centering
\hspace{-1cm}
{\epsfysize 8cm \epsffile{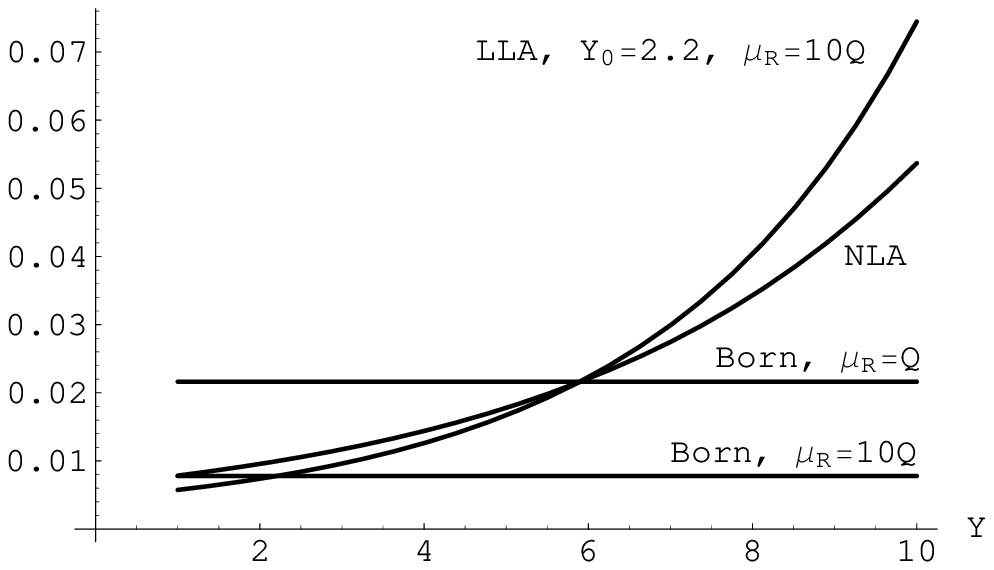}}
\caption[]{${\cal I}m_s ({\cal A}_{\mathrm series})Q^2/(s \, D_1 D_2)$ as a function of
$Y$ for optimal choice of the energy parameters $Y_0$ and $\mu_R$ (curve
labeled
by ``NLA''). The other curves represent the LLA result for $Y_0=2.2$ and $\mu_R=10Q$
and the Born (2-gluon exchange) limit for $\mu_R=Q$ and $\mu_R=10Q$.
The photon virtuality $Q^2$ has been fixed to 24 GeV$^2$ ($n_f=5$).}
\label{PMSres}
\end{figure}

These numbers make visible the effect of the NLA corrections: the $d_n$
coefficients are negative and increasingly large in absolute values as the
perturbative order increases. In such a situation it becomes necessary
to optimize the perturbative expansion, by proper choice of
the renormalization scale $\mu_R$ and of the energy scale $s_0$.

Several ways are known to optimize a perturbative expansion. In
Ref.~\cite{IP06} we adopted the principle of minimal sensitivity
(PMS)~\cite{Stevenson}. Usually PMS is used to fix the value of the
renormalization scale for the strong coupling. We used this principle
in a broader sense, requiring the minimal sensitivity of the predictions
to the change of both the renormalization and the energy scales,
$\mu_R$ and $s_0$. In Ref.~\cite{IP06} we considered the amplitude for
$Q^2$=24 GeV$^2$ and $n_f=5$ and studied its sensitivity to variation
of the parameters $\mu_R$ and $Y_0=\ln(s_0/Q^2)$. We could see that for
each value of $Y=\ln(s/Q^2)$ there are quite large regions in
$\mu_R$ and $Y_0$ where the amplitude is practically independent on $\mu_R$ and
$Y_0$ and we got for the amplitude a smooth behavior in $Y$, shown
in Fig.~\ref{PMSres}. The optimal values turned out to be $\mu_R\simeq 10 Q$
and $Y_0\simeq 2$, quite far from the kinematical values $\mu_R=Q$
and $Y_0=0$. These ``unnatural'' values are a manifestation
of the nature of the BFKL series: NLA corrections are large and then,
necessarily, since the exact amplitude should be renorm- and energy scale
invariant, the NNLA terms should be large and of the opposite sign with
respect to the NLA. These large NNLA corrections are mimicked by
the ``unnatural'' optimal values of $\mu_R$ and $Y_0$.

\begin{figure}[tb]
\centering
\hspace{-1cm}
{\epsfysize 8cm \epsffile{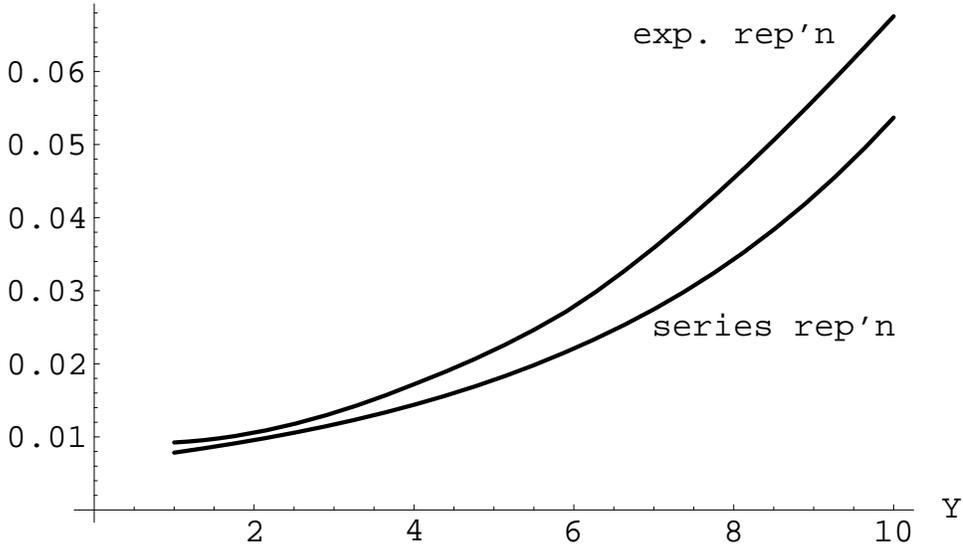}}
\caption[]{${\cal I}m_s ({\cal A})Q^2/(s \, D_1 D_2)$ as a function of $Y$
at $Q^2$=24 GeV$^2$ ($n_f=5$) from series and ``exponentiated'' representations,
in both cases with the PMS optimization method.}
\label{PMSexp_vs_PMSseries}
\end{figure}

As anticipated in the Introduction, it is important to have an estimate
of the systematic uncertainty which plagues our determination of the
energy behavior of the amplitude. The main sources of systematic effects
are given by the choice of the representation of the amplitude and by
the optimization method adopted. In the following, we compare the
determination of the amplitude at $Q^2$ = 24 GeV$^2$ ($n_f=5$)
through the PMS method, given in Fig.~\ref{PMSres}, with other determinations
obtained changing either the representation of the amplitude or the
optimization method.

At first, we compare the series and the ``exponentiated'' determinations
using in both case the PMS method. The procedure we followed to determine
the energy behavior of the ``exponentiated'' amplitude is straightforward:
for each fixed value of $Y$ we determined the optimal choice of the
parameters $\mu_R$ and $Y_0$ for which the amplitude given in
Eq.~(\ref{amplnlaE}) is the least sensitive to their variation. Also
in this case we could see wide regions of stability of the amplitude
in the $(\mu_R,Y_0)$ plane. The optimal values of $\mu_R$ and $Y_0$
are quite similar to those obtained in the case of the series representation,
with only a slight decrease of the optimal $\mu_R$. In
Fig.~\ref{PMSexp_vs_PMSseries} we show the result and compare it to
the PMS determination from the series representation. The two curves are
in good agreement at the lower energies, the deviation increasing for large
values of $Y$. It should be stressed, however, that the applicability
domain of the BFKL approach is determined by the condition
$\bar \alpha_s(\mu_R) Y \sim 1$ and, for $Q^2$=24 GeV$^2$ and for the
typical optimal values of $\mu_R$, one gets from this condition $Y\sim 5$.
Around this value the discrepancy between the two determinations is within a
few percent.

\begin{figure}[tb]
\centering
\hspace{-1cm}
{\epsfysize 8cm \epsffile{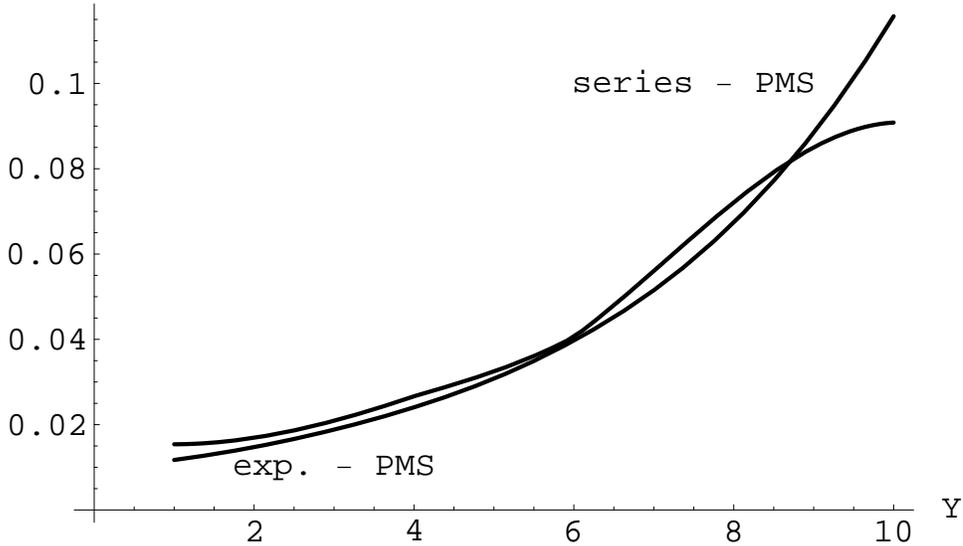}}
\caption[]{The same as Fig.~\ref{PMSexp_vs_PMSseries} at $Q^2$=5 GeV$^2$ ($n_f=4$).}
\label{PMSexp_vs_PMSseries_Q2=5}
\end{figure}

We repeated the same analysis at $Q^2$ = 5 GeV$^2$ ($n_f=4$) and compared the
PMS determination from the ``exponentiated'' amplitude with the PMS determination
from the series representation, obtained first in our paper Ref.~\cite{IP06}.
Fig.~\ref{PMSexp_vs_PMSseries_Q2=5} shows that the two determinations are in nice
agreement.
Despite that one should stress that in a case $Q^2$ = 5 GeV$^2$ we found 
for exponentiated amplitude much higher value for optimal energy scale,
$Y_0\simeq 6$. This may be an indication that the convergence of NLA BFKL
approach is actually worse for this smaller scale than it is for $Q^2$ = 24
GeV$^2$.

As a second check, we changed the optimization method and applied
it both to the series and to the ``exponentiated'' representation.
The method considered is the fast apparent convergence (FAC)
method~\cite{Grun}, whose strategy, when applied to a usual perturbative
expansion, is to fix the renormalization scale to the value for which the
highest order correction term is exactly zero. In our case, the application
of the FAC method requires an adaptation, for two reasons: the first is that
we have two energy parameters in the game, $\mu_R$ and $Y_0$, the second
is that, if only strict NLA corrections are taken, the amplitude
does not depend at all on these parameters.

\begin{figure}[tb]
\centering
\hspace{-1cm}
{\epsfysize 8cm \epsffile{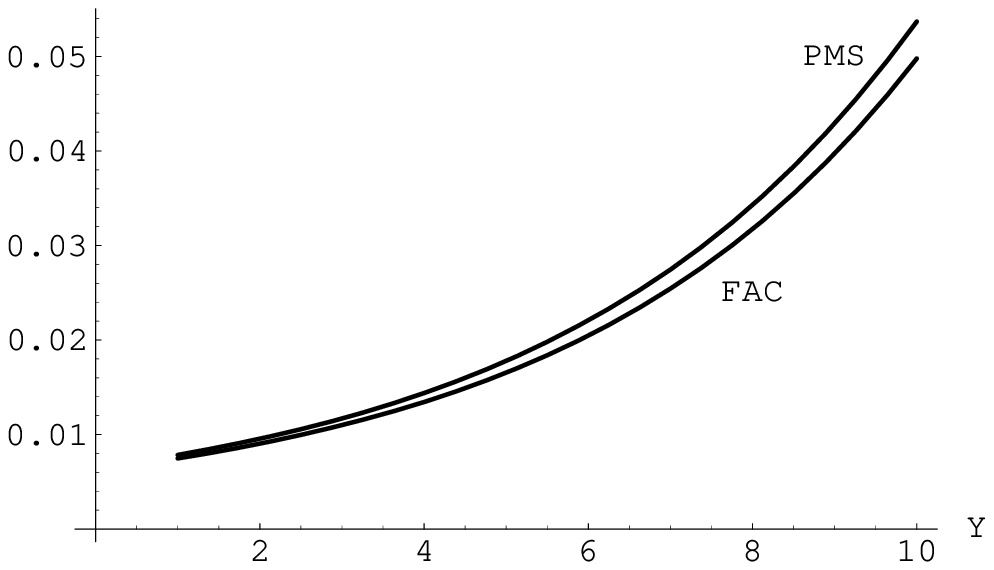}}
\caption[]{${\cal I}m_s ({\cal A}_{\mathrm series})Q^2/(s \, D_1 D_2)$ as a function of
$Y$ at $Q^2$=24 GeV$^2$ ($n_f=5$) from the series representation with PMS
and FAC optimization methods.}
\label{FAC_vs_PMS_series}
\end{figure}

Therefore, in the case of the series representation, Eq.~(\ref{series}), we
choose to put to zero the sum
\[
\frac{1}{(2\pi)^2}  \alpha_s(\mu_R)^2
\sum_{n=1}^{\infty}\bar \alpha_s(\mu_R)^n   \, b_n \,
d_n(s_0,\mu_R)\ln\left(\frac{s}{s_0}\right)^{n-1}
\]
and found for each fixed $Y$ the values of $\mu_R$ and $Y_0$ for which
the vanishing occurs. This gives a line of values in the $(\mu_R,Y_0)$ plane, among
which the optimal choice is done applying a minimum sensitivity criterion.
The result is shown in Fig.~\ref{FAC_vs_PMS_series}. The agreement with
the series representation with the PMS method is rather good over a wide energy range.

\begin{figure}[tb]
\centering
\hspace{-1cm}
{\epsfysize 8cm \epsffile{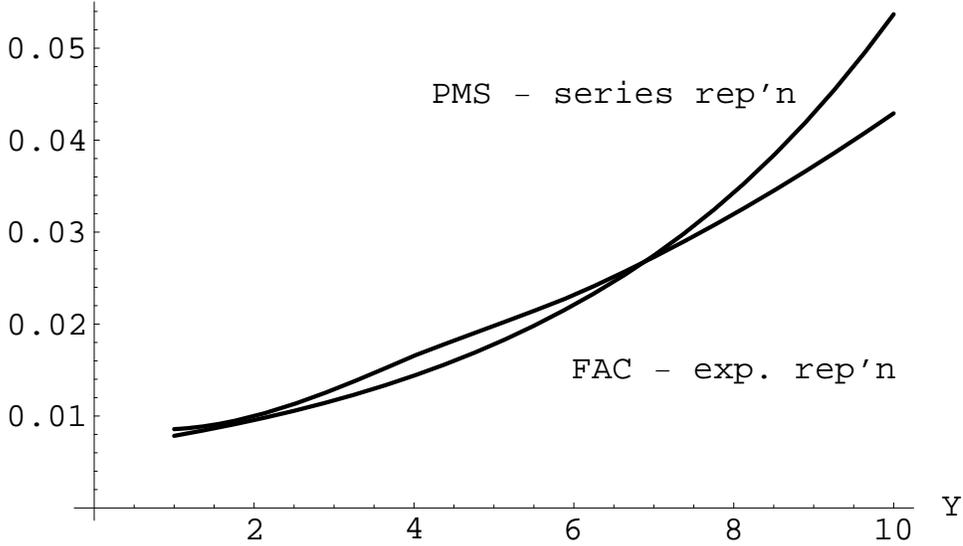}}
\caption[]{${\cal I}m_s ({\cal A})Q^2/(s \, D_1 D_2)$ as a function of
$Y$ at $Q^2$=24 GeV$^2$ ($n_f=5$) from the series representation with the PMS
optimization method and from the ``exponentiated'' representation with the FAC
optimization method.}
\label{FACexp_vs_PMSseries}
\end{figure}

In the case of the ``exponentiated'' amplitude'', Eq.~(\ref{amplnlaE}),
we proceeded in the same way, but requiring the vanishing of the
expression given by the R.H.S. of Eq.~(\ref{amplnlaE}) minus
the LLA amplitude, i.e.
\[
\frac{{\cal I}m_s\left( {\cal A}_{\mathrm exp} \right)}{D_1D_2}
-\frac{s}{(2\pi)^2}
\int\limits^{+\infty}_{-\infty}
d\nu \left(\frac{s}{s_0}\right)^{\bar \alpha_s(\mu_R) \chi(\nu)}
\alpha_s^2(\mu_R) c_1(\nu)c_2(\nu)\;.
\]
In Fig.~\ref{FACexp_vs_PMSseries} the result is compared with series
representation in the PMS method: there is nice agreement over the
whole energy range considered.

\begin{figure}[tb]
\centering
\hspace{-1cm}
{\epsfysize 8cm \epsffile{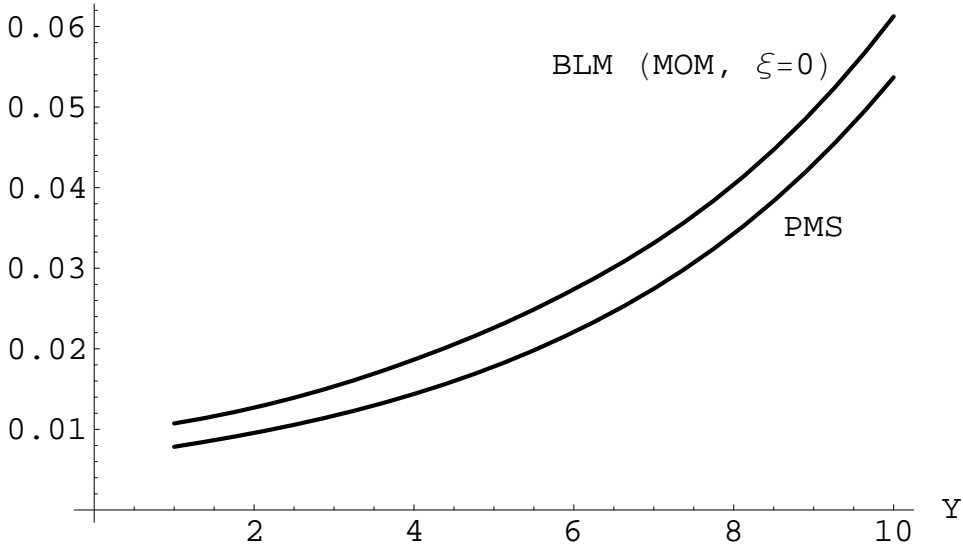}}
\caption[]{${\cal I}m_s ({\cal A})Q^2/(s \, D_1 D_2)$ as a function of
$Y$ at $Q^2$=24 GeV$^2$ ($n_f=5$) from the series representation with PMS
and BLM optimization methods.}
\label{BLM_vs_PMS_series}
\end{figure}

Another popular optimization procedure is the Brodsky-Lepage-Mackenzie (BLM)
method~\cite{BLM}, which amounts to perform a finite renormalization
to a physical scheme and then choose the renormalization scale in order to
remove the $\beta_0$-dependent part. We applied this method only to the
series representation, Eq.~(\ref{series}), and proceeded as follows: we
first performed a finite renormalization to the momentum (MOM) scheme
with $\xi=0$ (see Ref.~\cite{KIM1}),
\[
\alpha_s \to \alpha_s \left[1+T_{MOM}(\xi=0) \frac{\alpha_s}{\pi}\right]\;,
\hspace{1cm}
T_{MOM}(\xi=0)=T_{MOM}^{conf}+T_{MOM}^\beta \;,
\]
\[
T_{MOM}^{conf}=\frac{N_c}{8}\frac{17}{2} I \;, \hspace{1cm}
T_{MOM}^\beta=-\frac{\beta_0}{2}\left[1+\frac{2}{3}I\right]\;,
\hspace{1cm} I\simeq 2.3439 \;,
\]
then, we chose $Y_0$ and $\mu_R$ in order to make the term proportional
to $\beta_0$ in the resulting amplitude vanish. We observe that the
$\beta_0$-dependence in the series representation of the amplitude is hidden
into the $d_n$ coefficients, Eq.~(\ref{ds}). Among the resulting pairs of
values for $Y_0$ and $\mu_R$, we determined the optimal one according to
minimum sensitivity. This method has a drawback in our case, since for
each fixed $Y$, the optimal choice for $Y_0$ turned to be always
$Y_0\simeq Y$. However, if one blindly applies the procedure above, one
gets a curve which slightly overshoots the one for the series
representation with the PMS method, see Fig.~\ref{BLM_vs_PMS_series}.

\begin{figure}[tb]
\centering
\hspace{-1cm}
{\epsfysize 8cm \epsffile{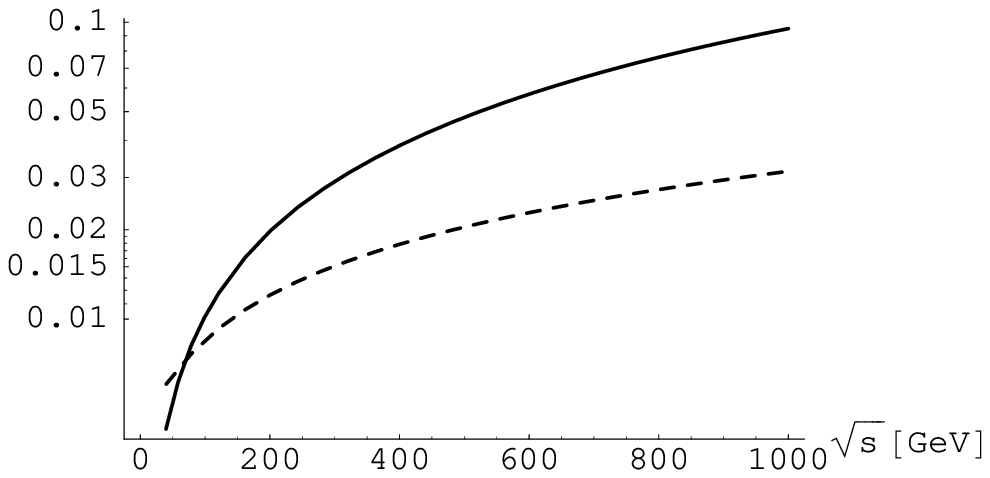}}
\caption[]{Linear-log plot of $\left. d\sigma/dt\right|_{t=t_0}$ [pb/GeV$^2$] as a function of
$\sqrt{s}$ at $Q^2$=16 GeV$^2$ ($n_f=4$) from the series representation with the PMS
optimization method (solid line) compared with the determination from the approach in
Ref.~\cite{EPSW1} (dashed line).}
\label{dsig_Q2_16}
\end{figure}

\begin{figure}[tb]
\centering
\hspace{-1cm}
{\epsfysize 8cm \epsffile{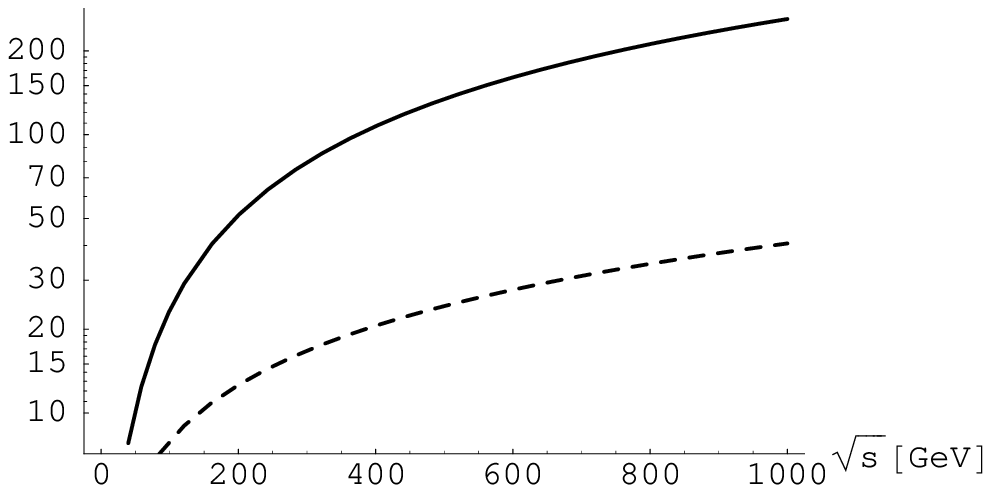}}
\caption[]{Linear-log plot of $\left. d\sigma/dt\right|_{t=t_0}$ [pb/GeV$^2$] as a function of
$\sqrt{s}$ at $Q^2$=4 GeV$^2$ ($n_f=3$) from the series representation with the PMS
optimization method (solid line) compared with the determination from the approach in
Ref.~\cite{EPSW1} (dashed line).}
\label{dsig_Q2_4}
\end{figure}

\section{The differential cross section at the minimum $|t|$:
comparison with a different approach}

The $\gamma^* \gamma^* \to \rho \rho $ process at the lowest order
(two-gluon exchange in the $t$-channel) was studied in Ref.~\cite{Pire:2005ic}.
At that level our results coincide, see also~\cite{IP06}.
The same process with the inclusion of NLA BFKL effects has been considered
in Ref.~\cite{EPSW1}. In that paper,
the amplitude has been built with the following ingredients: leading-order
impact factors for the $\gamma^* \to \rho$ transition, BLM scale fixing
for the running of the coupling in the prefactor of the amplitude (the BLM
scale is found using the NLA $\gamma^* \to \rho$ impact factor
calculated in Ref.~\cite{IKP04}) and renormalization-group-resummed
BFKL kernel, with resummation performed on the LLA BFKL kernel at fixed
coupling~\cite{KMRS04}. In Ref.~\cite{EPSW1} the behavior of $d\sigma/dt$ at $t=t_0$
was determined as a function of $\sqrt{s}$ for three values of
the common photon virtuality, $Q$=2, 3 and 4 GeV.

In order to make a comparison with the findings of Ref.~\cite{EPSW1},
we computed $d\sigma/dt$ at $t=t_0$ for $Q$=2 and $Q$=4 GeV as functions
of $\sqrt{s}$. We used $f_\rho$=216 MeV, $\alpha_{\mathrm EM}=1/137$ and
the two--loop running strong coupling corresponding to the value
$\alpha_s(M_Z)=0.12$. The results are shown in the linear-log
plots of Figs.~\ref{dsig_Q2_16} and \ref{dsig_Q2_4}, which show a large
disagreement. It would be interesting to understand to what extent
this disagreement is due to the use in Ref.~\cite{EPSW1} of
LLA impact factors instead of the NLA ones or to the way the collinear
improvement of the kernel is performed.

\begin{figure}[tb]
\centering
\hspace{-1cm}
{\epsfysize 8cm \epsffile{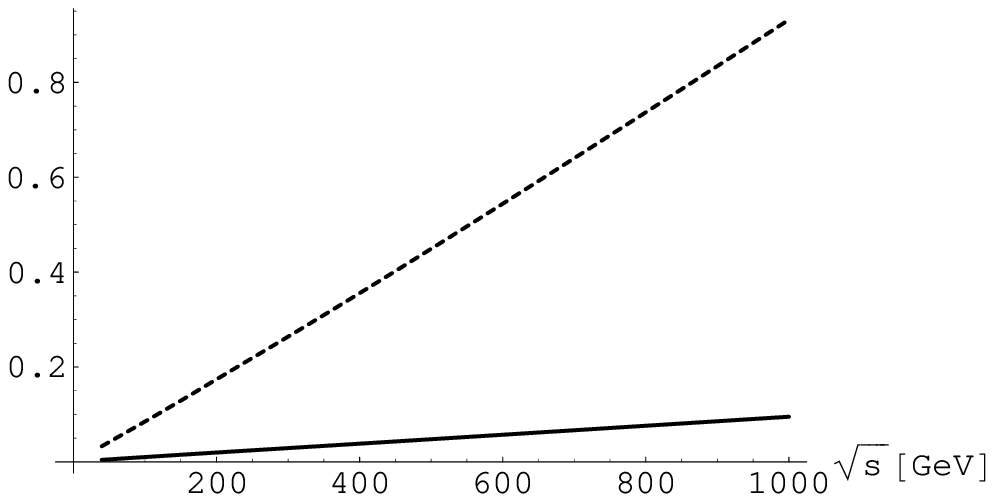}}
\caption[]{Linear plot of $\left. d\sigma/dt\right|_{t=t_0}$ [pb/GeV$^2$] as a function of
$\sqrt{s}$ at $Q^2$=16 GeV$^2$ ($n_f=4$) from the series representation with the PMS
optimization method using NLO impact factors (solid line) and LO impact factors (dashed line).}
\label{dsig_Q2_16_LOvsNLO}
\end{figure}

\begin{figure}[tb]
\centering
\hspace{-1cm}
{\epsfysize 8cm \epsffile{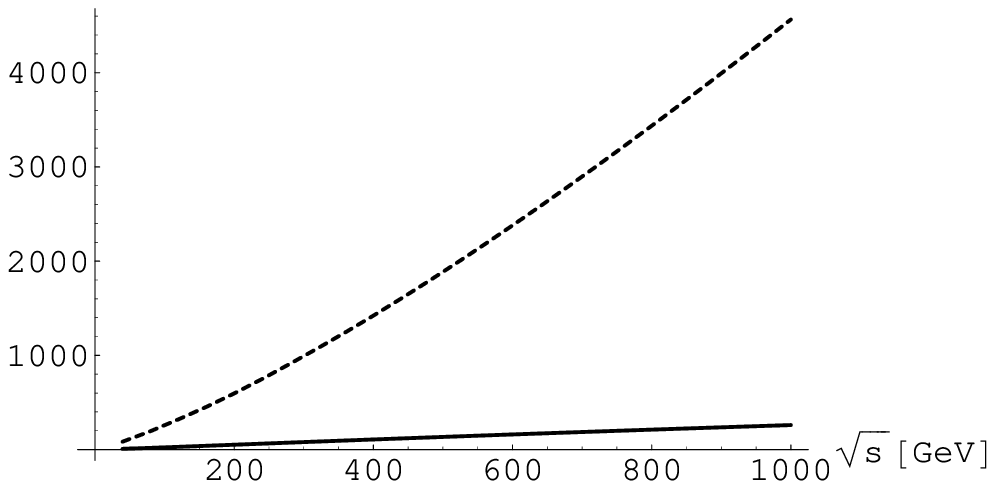}}
\caption[]{Linear plot of $\left. d\sigma/dt\right|_{t=t_0}$ [pb/GeV$^2$] as a function of
$\sqrt{s}$ at $Q^2$=4 GeV$^2$ ($n_f=3$) from the series representation with the PMS
optimization method (solid line) compared with the determination from the approach in
Ref.~\cite{EPSW1} (dashed line).}
\label{dsig_Q2_4_LOvsNLO}
\end{figure}

In order to understand to what extent the discrepancy is due to the use of leading order (LO)
impact factors instead of next-to-leading order (NLO) ones, we repeated our determination of
$d\sigma/dt$ at $t=t_0$ for $Q$=2 and $Q$=4 GeV, using LO impact factors and keeping
from the their NLO contribution only the terms proportional to $\ln[s_0/(Q_1 Q_2)]$
and to $\ln[\mu_R^2/(Q_1 Q_2)]$ which are universal and needed to guarantee the $s_0$-
and $\mu_R$-independence of the amplitude with NLA accuracy. The result is that
$d\sigma/dt$ at $t=t_0$ increases roughly by an order of magnitude with respect to
our previous determination (see Figs.~\ref{dsig_Q2_16_LOvsNLO} 
and~\ref{dsig_Q2_4_LOvsNLO})
and therefore the disagreement with~\cite{EPSW1} becomes even worse. This is not
surprising: impact factors give a sizable contribution to the NLA part of the
amplitude which is negative with respect to the LLA part; if they are kept at LO,
the NLA part of the amplitude is less negative and the total amplitude is therefore
increased.

\section{Conclusions}

We have studied the amplitude for the forward transition from two virtual
photons with equal virtuality to two light vector mesons in the Regge limit of
QCD with next-to-leading order accuracy. We have found that its behavior
with the center-of-mass energy is stable in the applicability region of
the BFKL approach under change of representation of the amplitude and
under change of the method of optimization of the perturbative series.

We have determined also the differential cross section at the minimum $|t|$
for two values of the common photon virtuality and found strong disagreement
with another determination based on the inclusion of next-to-leading order effects
only through the kernel and on the use of collinear improvement of the kernel
at the leading order.

\section{Acknowledgments}

We thank V.S. Fadin and A. Sabio Vera for many stimulating discussions.
D.I. thanks the Dipartimento di Fisica dell'Universit\`a della Calabria
and the Istituto Nazionale di Fisica Nucleare (INFN), Gruppo collegato di Cosenza,
for the warm hospitality while part of this work was done and for the
financial support. The work of D.I. was also partially supported by
grants RFBR-05-02-16211, NSh-5362.2006.2.

\end{document}